\documentclass{article}
\usepackage{spconf,amsmath,graphicx,hyperref}
\usepackage{booktabs}
\usepackage{threeparttable}
\usepackage{xcolor}
\usepackage[hypcap=true]{caption}
\usepackage{microtype}
\usepackage{algorithm,algpseudocode}
\usepackage{mathtools}
\definecolor{linkcolor}{RGB}{83,83,182}
\hypersetup{
  colorlinks=true,
  citecolor=linkcolor,
  linkcolor=linkcolor
}
\usepackage[nameinlink,capitalize]{cleveref}

\Crefname{equation}{Eq.}{Eqs.}
\Crefname{figure}{Fig.}{Figs.}
\Crefname{table}{Tab.}{Tabs.}
\Crefname{section}{Sec.}{Secs.}
\Crefname{algorithm}{Alg.}{Algs.}
\renewcommand{\Cref}[1]{\cref{#1}}

\graphicspath{{images/}}

\title{Adapting offline models to a streaming context for music source separation}
\name{Dylan Sechet, Marc Evrard, Matthieu Kowalski\thanks{Code and audio samples: \url{https://dylansechet.com/papers/mss-streaming}}
}
\address{
  Université Paris-Saclay, Inria, CNRS \\
  Laboratoire Interdisciplinaire des Sciences du Numérique \\
  \texttt{firstname.lastname@lisn.fr}
}
\begin{document}
\ninept
\maketitle
\begin{abstract}
Real-time music source separation must satisfy two constraints: a bound on algorithmic latency and a bound on computational cost. Offline separators are usually omitted from real-time comparisons or credited with a latency equal to their full input length. We show that this latency is set by where the output is read, not by the length of the separator's input. An unmodified offline model can therefore run in a streaming setting, without retraining. At each step, the input slides by one STFT hop, and one output hop is read out. The resulting latency can be as low as one STFT hop ($23$~ms), and the computational cost does not increase as latency shrinks.
We identify a theoretical model-dependent latency boundary below which separation quality should drop steeply, and confirm this experimentally across three architectures. 
At equal algorithmic latency, streamed off-the-shelf checkpoints for HT-Demucs and SCNet match the published results of dedicated real-time models in terms of separation quality. Streamed models remain far less computationally efficient: only HT-Demucs runs faster than real time on a GPU.
% \looseness=-1
\end{abstract}
\begin{keywords}
  Music source separation, real-time source separation, streaming inference, algorithmic latency, low-latency audio processing
\end{keywords}

\section{Introduction}\label{sec:intro}
Music source separation aims to decompose a musical mixture into its constituent sources, traditionally vocals, drums, bass, and others.
Deep learning has driven rapid progress, and the strongest current systems, based on large neural networks, separate these sources with high fidelity \cite{luMusicSourceSeparation2024, tongSCNetSparseCompression2024, rouardHybridTransformersMusic2022}.
Most systems are designed for offline use: they process the entire recording, exploiting both past and future context.

Deploying a separation model in real time imposes two independent constraints \cite{venkateshRealTimeLowLatencyMusic2024, wuPracticalRealTimeLowLatency2025}.
The first is algorithmic latency, the theoretical delay inherent in the algorithm's design, independent of hardware performance or available computing power. The second is computational latency, the processing time needed to run the model, which depends on the available computational power.
Dedicated real-time models are designed to satisfy both at once. For algorithmic latency, they tend to use small STFT windows, reaching latencies as low as $23$~ms \cite{venkateshRealTimeLowLatencyMusic2024, kimContrastiveLearningBased2023}, which suits applications with strict latency budgets. In terms of computational cost, they are lightweight and therefore both fast and usable on resource-constrained devices. Band-SCNet, for instance, reports a real-time factor of $0.48$ on a single CPU thread with $2.59$~M parameters \cite{yangBandSCNetCausalLightweight2025}.
However, meeting both constraints incurs a cost to separation quality: real-time models still trail standard offline models.

These offline models are hard to compare with dedicated real-time models, especially when measuring their latency. In practice, they are often left out of real-time comparisons. When included, their algorithmic latency is tabulated as their full input length, between $2$ and $15$~s \cite{venkateshRealTimeLowLatencyMusic2024, wuPracticalRealTimeLowLatency2025}, or they are placed in a separate non-real-time category~\cite{yangBandSCNetCausalLightweight2025}.
We argue that the algorithmic latency of an offline model is determined by its output readout configuration, not by its input length: reading out a single STFT hop from the model's output reduces latency to that hop length, typically $23$~ms, without requiring modifications to the model or its STFT pipeline. 
Offline models then match algorithmic latency with the fastest dedicated real-time models: a latency equal to the input length is only one possible operating point, and it is roughly two orders of magnitude higher than the minimum achievable latency.

In this work, we stream off-the-shelf offline models at selected algorithmic latencies ranging from $23.2$~ms to $3$~s and measure the impact of this choice on separation quality. The lower end of this latency range matches that of dedicated real-time models, while latencies around $100$~ms remain sufficient for less latency-critical uses such as karaoke or remixing of broadcast music. We make three contributions:

\begin{itemize}
  \item \textbf{Streaming offline models} (\Cref{sec:streaming}). We show that the algorithmic latency of a streamed offline separator is determined by its readout point, not by its input length: reading out a single hop of the model output reduces it to $23.2$~ms. Lookahead then trades latency for quality at a fixed computational cost, with two regimes on either side of a latency equal to the length of a single STFT window. The analysis predicts that quality starts to drop steeply for latencies below each model's STFT window length.
  \item \textbf{Comparison with dedicated models} (\Cref{sec:results}). Streamed off-the-shelf checkpoints of HT-Demucs and SCNet, used without retraining, match the best published dedicated real-time models at $23$ and $93$~ms. All streamed models remain far less efficient; only HT-Demucs runs faster than real time on a GPU.
  \looseness=-1
  \item \textbf{Cost of the streaming regime} (\Cref{sec:results}). We quantify how separation quality depends on future context across architectures and sources, and show through an ablation that fine-tuning with a streaming-specific loss does not reduce this dependency.
\end{itemize}

\section{Related Work}\label{sec:related}
%
%We first review dedicated real-time separation models, then the closest prior work, which streams offline models.
%
In music source separation, dedicated real-time models have received little attention compared with the offline setting \cite{venkateshRealTimeLowLatencyMusic2024, wuPracticalRealTimeLowLatency2025}.
This has not prevented multiple real-time models from being proposed, usually by adapting existing offline architectures.
Deep Latent Masking \cite{kimContrastiveLearningBased2023} operates directly in the time domain, at an algorithmic latency of $23$~ms. It estimates a mask for each source in the latent space of a U-Net with a dual-path RNN bottleneck.
HS-TasNet \cite{venkateshRealTimeLowLatencyMusic2024} introduces spectral processing: it draws on Hybrid Demucs \cite{defossezHybridSpectrogramWaveform2022}, combining a waveform branch and a spectrogram branch with unidirectional LSTMs, also at $23$~ms.
RT-STT \cite{wuPracticalRealTimeLowLatency2025} bases its architecture on a more modern spectrogram-based U-Net, DTTNet \cite{chenMusicSourceSeparation2024}, with more than ten times fewer parameters. It also reaches a latency of $23$~ms and shows how quantization can reduce computational cost.
Finally, both Online SCNet and Band-SCNet \cite{yangBandSCNetCausalLightweight2025} are built by making SCNet \cite{tongSCNetSparseCompression2024} causal. Online SCNet naively replaces its non-causal layers with causal counterparts, and Band-SCNet then adds cross-band and narrow-band blocks to recover part of the lost separation performance. Both reach an algorithmic latency of $93$~ms.

While real-time methods remain marginal in music source separation, they are well studied in speech processing, where data challenges impose explicit latency budgets \cite{reddyICASSP2021Deep2021}.
Défossez et al.\ make the Demucs architecture causal with a unidirectional LSTM and train it for speech enhancement, reaching faster-than-real-time operation on a single laptop CPU core \cite{defossezRealTimeSpeech2020}.
Research in speech processing also measures how much future context, or lookahead, a streaming model needs: its impact on enhancement is negligible for Wilson et al.\ \cite{wilsonExploringTradeoffsModels2018}, a single training-time hyperparameter sets it in LaCo-SENet \cite{kimLatencyConfigurableStreamingSpeech2026b}, and a speech recognizer chooses it at runtime in~\cite{strimelLookaheadWhenIt2023b}.
% \looseness=-1
The aforementioned models are all designed or trained for streaming. Morrone et al.\ instead stream an offline speech separator and study the effect of lookahead  \cite{morroneConversationalSpeechSeparation2022}. 
They advance their input by $0.5\text{ s}$ per processing pass and emit a corresponding $0.5\text{ s}$ output, establishing a latency floor of $0.5\text{ s}$ below which they experience substantial performance degradation.
In contrast, advancing by a single STFT hop lowers the latency floor to $23\text{ ms}$, enabling operation at sub-window latencies, a regime their setup never enters. 
For music, we find that streaming at one window ($93$~ms) costs less than $1$~dB relative to offline setup performance, and that quality drops steeply only below that threshold.

\section{Streaming offline models}\label{sec:streaming}
An offline separator assumes that the entire recording is available and processes it in fixed-length segments, which we call the model's \textit{context window}.
To stream such a model, we slide its context window over the incoming mixture, one STFT hop at a time.

Let $C$ denote the length of the context window in samples, and let $W$ and $H$ denote the window length and hop size of the model's STFT. 
At each step, the model receives the $C$ most recent mixture samples and returns $C$ samples per source. Emitting this whole output would incur a latency of $C$. Instead, we emit a single buffer of $H$ samples, whose position determines how much future context the model has seen when estimating them.

\subsection{Output readout and latency}

\begin{figure}[h]
  \centering
  \includegraphics[width=0.3\columnwidth]{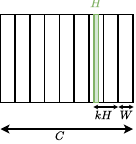}
  \caption{Readout position in a model output of size $C$: the pipeline emits the green segment, with latency $W+kH$.}
  \label{fig:readout}
\end{figure}

We parameterize the readout position by an integer lookahead $k$, measured in STFT hops and possibly negative, and emit the output samples $[n_k,n_k+H)$, with $n_k=C-W-kH$.
The corresponding algorithmic latency is
\begin{equation}
L(k) = W+kH.
\label{eq:latency}
\end{equation}
Each unit of $k$ thus adds one hop of latency (see \Cref{fig:readout}). 
The emitted hop must remain inside the model output, $n_k+H\leq C$, which bounds the lookahead and the latency:
\begin{equation}
k \geq 1-\frac{W}{H}, \qquad L_{\min}=W + \Big(1-\frac{W}{H}\Big) H =H.
\end{equation}
This bound follows from our readout construction as a buffer of size $H$, and is not a structural lower bound for streaming. At $f_s=44.1$~kHz and $H=1024$, $L_{\min}=23.2$~ms for all models considered here, while $L(0)=W$ is $92.9$~ms for HT-Demucs and SCNet ($W=4096$) and $139.3$~ms for DTTNet ($W=6144$).

\begin{figure}[h]
  \centering
  \includegraphics[width=\columnwidth]{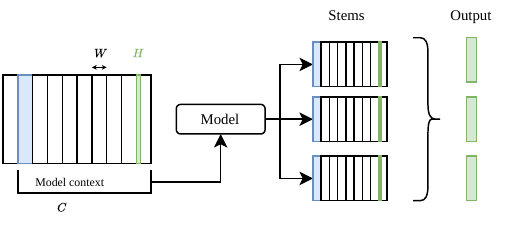}
  \caption{Streaming with lookahead $k=0$ and latency $W$. The oldest window in the model's context is in blue.}
  \label{fig:pipeline}
\end{figure}

The output buffer corresponds to the overlap-add of $W/H$ inverse STFT synthesis frames (or the spectral branch for HT-Demucs). The configuration $k = 0$, illustrated in \Cref{fig:pipeline}, separates two regimes.

For $k\geq0$, all these frames lie within the observed input, and each hop of additional lookahead adds context only to the right.
For $k<0$, $|k|$ synthesis frames extend past the last observed sample and are completed by the model's own STFT padding (reflection for all models considered here). 
Below a latency of $W$, the lack of future context forces part of the emitted hop's synthesis to rely on padding instead of real input.
Padding is thus absent for a latency $L(k) \geq W$. It grows with $|k|$ for $L(k)<W$, implying a characteristic knee in separation quality at $L = W$. Note that this latency threshold is not fixed, but rather determined by each model's STFT configuration.

\subsection{Streaming procedure}

\begin{algorithm}[h]
\caption{Streaming an offline separator $f$ with lookahead $k$.}
\label{alg:streaming}
\begin{algorithmic}[1]
\Require pretrained separator $f$, lookahead $k \geq 1-W/H$
\State $n_k \gets C-W-kH$
\For{each incoming block $\mathbf{x}_t$ of $H$ samples}
  \State $\mathbf{b} \gets [\,\mathbf{b}_{H:C},\ \mathbf{x}_t\,]$ \Comment{slide by one hop}
  \State $\hat{\mathbf{s}} \gets f(\mathbf{b})$ \Comment{$C$ samples per source}
  \State \textbf{emit} $\hat{\mathbf{s}}_{n_k:n_k+H}$ \Comment{latency $W+kH$}
\EndFor
\end{algorithmic}
\end{algorithm}

\Cref{alg:streaming} summarizes the procedure.
For $k\geq0$, the emitted hop is synthesized from the same STFT frames as in the offline setting, although its estimate still differs from the offline configuration: the network's context now ends $L(k)$ samples after the hop.
For $k<0$, the estimate is equivalent to the offline model when deprived of future context. 
%is not that of a causal separator, but that of the unchanged offline model deprived of future input.
In both regimes, a single hop is produced during each forward pass, so streaming cannot use the overlap-add cross-fading that offline pipelines typically apply between chunks.
% \looseness=-1

The choice of $k$ does not affect the model input or the number of forward passes, hence the computational cost does not depend on algorithmic latency.
Streaming is nevertheless less efficient than offline processing: each pass computes $C$ output samples but keeps only $H$, discarding most of the model's output at every step.

\section{Experimental Setup}\label{sec:setup}
We evaluate models on the standard MUSDB18-HQ dataset \cite{rafiiMUSDB18HQUncompressedVersion2019}, reporting the Signal-to-Distortion Ratio (SDR) \cite{vincentPerformanceMeasurementBlind2006} as computed by \texttt{museval} \cite{stoter2018SignalSeparation2018}. 
We focus on mid-sized models rather than the largest ones ($\geq 50$~M parameters), such as BS-RoFormer \cite{luMusicSourceSeparation2024} and Band-Split RNN \cite{luoMusicSourceSeparation2023}: recomputing the full context at every hop already brings the models we use to a real-time factor (RTF) near 1.
Our comparison includes DTTNet \cite{chenMusicSourceSeparation2024}, a TFC-TDF U-Net with an RNN bottleneck; HT-Demucs \cite{rouardHybridTransformersMusic2022}, whose cross-domain transformer combines waveform and spectrogram information; and SCNet \cite{tongSCNetSparseCompression2024}, a recurrent architecture built on band-split representations.
% \looseness=-1

\begin{table*}[ht]
  \centering
  \caption{Median SDR (dB) on MUSDB18-HQ, with rows grouped by algorithmic latency.}
  \label{tab:main}
  \begin{threeparttable}
    \begin{tabular}{llcr@{~}lccccc}
      \toprule
      Model & Params (M) & Lat. (ms) & \multicolumn{1}{c}{All} & & Drums & Bass & Other & Vocals & RTF \\
      \midrule
      HS-TasNet \cite{venkateshRealTimeLowLatencyMusic2024}\tnote{$\dagger$}  & 42.0 & 23 & 4.65 & & 5.22 & 4.59 & 3.64 & 5.13 & - \\
      RT-STT \cite{wuPracticalRealTimeLowLatency2025}\tnote{$\dagger$}  & 0.38 & 23 & 5.17 & & 5.83 & 5.25 & 4.02 & 5.56 & - \\
      Deep Latent Masking \cite{kimContrastiveLearningBased2023}\tnote{$\dagger\ddagger$} & 5.7  & 23 & 6.47 & & 7.05 & 7.29 & 4.62 & 6.91 & - \\
      HT-Demucs streamed (ours)\tnote{$\S$ } & 42.0  & 23 & 6.97 & [6.29, 7.34] & \textbf{8.17} & \textbf{7.87} & 4.95 & 6.90 & 0.89 \\
      SCNet streamed (ours)\tnote{$\S$} & 10.08 & 23 & \textbf{6.99} & [6.50, 7.36] & 7.65 & 7.20 & \textbf{5.66} & 7.45 & 1.59 \\
      DTTNet streamed (ours)       & 5.0 ($\times 4$) & 23 & 6.06 & [5.42, 6.36] & 5.67 & 5.44 & 5.05 & \textbf{8.10} & 1.57 \\
      \midrule
      Online SCNet \cite{yangBandSCNetCausalLightweight2025}\tnote{$\dagger$}  & 4.36 & 93 & 7.14 & & 8.23 & 6.16 & 5.64 & 8.53 & - \\
      Band-SCNet \cite{yangBandSCNetCausalLightweight2025}\tnote{$\dagger$}    & 2.59 & 93 & 7.79 & & 9.44 & 7.13 & 5.87 & 8.74 & - \\
      HT-Demucs streamed (ours)\tnote{$\S$} & 42.0  & 93 & 8.47 & [7.64, 9.07] & \textbf{9.74} & \textbf{9.43} & 6.20 & 8.49 & 0.89 \\
      SCNet streamed (ours)\tnote{$\S$} & 10.08 & 93 & \textbf{8.51} & [7.73, 9.07] & 9.69 & 9.17 & \textbf{6.49} & 8.67 & 1.59 \\
      DTTNet streamed (ours)       & 5.0 ($\times 4$) & 93 & 7.51 & [6.62, 8.10] & 7.07 & 7.16 & 6.11 & \textbf{9.69} & 1.57 \\
      \midrule
      HT-Demucs\tnote{$\S$} & 42.0  & - & 8.84 & [7.90, 9.42] & 10.15 & 9.82 & 6.48 & 8.91  & 0.0053 \\
      SCNet\tnote{$\S$}            & 10.08 & - & \textbf{9.25} & [8.31, 9.78] & \textbf{10.28} & \textbf{10.15} & \textbf{7.21} & 9.35 & 0.0067 \\
      DTTNet                       & 5.0 ($\times 4$) & - & 7.94 & [7.08, 8.55] &  7.22 & 7.63 & 6.56 & \textbf{10.34} & 0.0123 \\
      \bottomrule
    \end{tabular}
    \begin{tablenotes}\footnotesize
    \item[$\dagger$] Results reported in the original papers. $\ddagger$ Evaluated on MUSDB18 (not HQ).
    \item[$\S$] Best off-the-shelf checkpoint, trained on MUSDB18-HQ plus $800$ songs (HT-Demucs) or MoisesDB~\cite{igorpereiraMoisesDBDatasetSource2023} (SCNet).
    \end{tablenotes}
  \end{threeparttable}
\end{table*}

We compare the fully offline pipeline (including overlap-add if necessary) with a version that is streamed following \Cref{sec:streaming}, to measure the streaming loss.
We then compare these streamed models at each level of algorithmic latency against the dedicated real-time architectures in \Cref{sec:related}.
All reported results use the pretrained weights released by their authors, without retraining or structural modification. \Cref{sec:fine_tune} fine-tunes two models only as an ablation.
% \looseness=-1

Following \Cref{sec:streaming}, the context window advances by one STFT hop, which corresponds to $1024$ samples for all three models.
We sweep the lookahead over its full range, from its lower bound $k = 1 - W/H$ to $k = 128$, giving latencies from $23.2$~ms to just over $3$~s.
% \looseness=-1

We also report the real-time factor (RTF), the ratio of the computation time to an audio's duration. An RTF below $1$ thus indicates real-time capability.
Execution times are measured on a single NVIDIA RTX 5080 GPU with a batch size of $1$ on models compiled with static input shapes, and we report the median of $50$ forward passes following $25$ warm-up iterations.
While the streamed model runs one forward pass for every hop, the offline overlap-add pipeline performs one pass per stride, which is half the context window at $50\%$ overlap. To obtain a per-sample processing time, we therefore normalize the forward-pass time by the hop size for streamed models and by the stride for offline baselines. 
DTTNet is an ensemble of four single-source models sharing the same architecture; we report the execution time of a single model, assuming all four models would be run in parallel in a production environment.
% \looseness=-1

\section{Results}\label{sec:results}

\begin{figure}[h]
  \centering
  \includegraphics[width=\columnwidth]{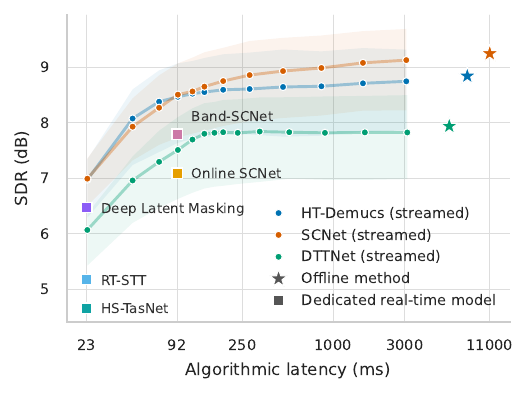}
  \caption{SDR against algorithmic latency (log scale) on MUSDB18-HQ. Each circle is one lookahead $k$ of a model's sweep, and shaded bands are 95\% confidence intervals. RTF is constant along each curve: $0.89$ (HT-Demucs), $1.59$ (SCNet), $1.57$ (DTTNet). Offline methods are placed at their context window length ($C$ in \cref{alg:streaming}).}
  \label{fig:latency_sdr}
\end{figure}

\Cref{tab:main} and \Cref{fig:latency_sdr} report SDR against algorithmic latency, with 95\% confidence intervals from bootstrapping \cite{efronIntroductionBootstrap1993, Confidence_Intervals}.

\subsection{Impact of streaming on model performance}
As expected, the separation quality of a streamed model rises with its lookahead. At $3$~s of algorithmic latency, all models come within $0.12$~dB of their offline equivalent.
As predicted in \Cref{sec:streaming}, quality starts to drop steeply below $k=0$ for all three models: below $93$~ms for HT-Demucs and SCNet, and below $139$~ms for DTTNet. The knee follows each model's STFT window rather than a common latency. HT-Demucs is also affected: its output sums both branches, and the padding in the spectral branch alone is sufficient to degrade it.
For $k \geq 0$, performance rises logarithmically with latency for HT-Demucs and SCNet, while remaining near flat for DTTNet, which does not appear to make much use of future context in its inference.
For $k<0$, padding replaces real input, and performance falls steeply. From $93$ to $23$~ms, all three models lose about $1.5$~dB, although DTTNet already operates at $k=-2$ at $93$~ms. The imperfect reconstruction for these negative lookaheads causes windowing artifacts. Perceptually, they are highly noticeable for up to half a window latency and manifest as a background buzz rather than clicks. 
% \looseness=-1

\begin{figure}[!h]
  \centering
  \includegraphics[width=\columnwidth]{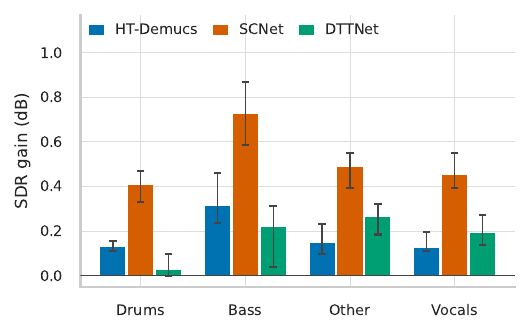}
  \caption{Per-source SDR gain from extending the lookahead from $k=0$ to $k=128$ hops, as the median of per-track SDR differences. Error bars are 95\% CIs from a paired per-track analysis.}
  \label{fig:lookahead_gain}
\end{figure}

\Cref{fig:lookahead_gain} breaks down the gain from future context by model and source. 
The model sets the magnitude of the gain, while the source sets part of its ordering: drums gain least for all three models. 
Bass gains the most with HT-Demucs and SCNet, plausibly because low-frequency components need longer context to be resolved.

\subsection{Effect of fine-tuning models for a streaming setting}\label{sec:fine_tune}
Comparing streamed SCNet with Online SCNet lets us break down the cost of adapting SCNet for real-time use.
Online SCNet makes SCNet causal (causal convolutions, unidirectional LSTM) and adds no compensation mechanism to recover the quality loss, so its gap to offline SCNet combines the loss from missing future context with the loss from the causal architecture itself. 
Our streamed checkpoint at the same latency pays only the future-context loss.
After correcting for training data (the additional MoisesDB training data adds $0.25$~dB \cite{tongSCNetSparseCompression2024}), Online SCNet trails offline SCNet in median SDR by $1.86$~dB, while streaming SCNet costs only $0.74$~dB.
At least $1.1$~dB of this gap therefore comes from the architectural changes and the associated reduction in parameter count, as losing future context costs at most $0.74$ dB.
The $0.74$~dB is an upper bound here, as that figure also absorbs the loss of overlap-add cross-fading.
% \looseness=-1

% The sweep further shows what the $0.74$~dB streaming cost is made of: at $k=128$, streamed SCNet has $3$~s of future context but still lacks cross-fading, and lies within $0.12$~dB of offline.
% Going from $k=0$ to $k=128$ thus recovers $0.62$~dB in median SDR, and the remaining streaming-cost effects amount to at most $0.12$~dB.

Whether this drop in performance at low lookaheads is avoidable may depend on the training objective: the model's training loss weights every predicted sample
equally, whereas streaming keeps only the samples at the readout. 
As an ablation, we fine-tuned SCNet and HT-Demucs with a loss restricted to the readout buffer, for both $k=0$ and $k=k_{\min} = -3$.
A control fine-tune using the model's original, unconstrained loss isolates the effect of further training from the effect of the restricted loss. 
Both runs use MUSDB18-HQ, $20$~epochs and a learning rate of $5\times 10^{-5}$. 
Relative to the control, the restricted loss gains at most $0.1$~dB for both lookaheads, suggesting that the loss from streaming is structural: this fine-tuning does not reduce it.
\looseness=-1

\subsection{Comparison to dedicated real-time architectures}
Streamed HT-Demucs and SCNet at least match the SDR of the best dedicated architectures of \Cref{tab:main} at both $23$ and $93$~ms. 
Their leads, $0.5$ to $0.7$~dB, are comparable to the width of our confidence intervals. 
DTTNet, on the other hand, trails Deep Latent Masking by $0.41$~dB at $23$~ms and Band-SCNet by $0.28$~dB at $93$~ms.

At $93$~ms, streamed SCNet leads Band-SCNet by $0.72$~dB.
At $23$~ms, both lead Deep Latent Masking by about $0.5$~dB, although the latter was evaluated on MUSDB18 rather than MUSDB18-HQ.

These comparisons rely on the best available models, which are not necessarily trained on the same datasets. 
The extra $800$ songs of the HT-Demucs checkpoint are worth about $1.5$~dB offline~\cite{rouardHybridTransformersMusic2022}, which is more than its lead, but the additional training data from MoisesDB adds only $0.25$~dB to SCNet~\cite{tongSCNetSparseCompression2024}, well below its lead.
Our streaming procedure does not require retraining or accessing the training data, so a stronger public checkpoint can be used as is. 

Only HT-Demucs runs faster than real time on our GPU, with a median RTF of $0.89$. The comparison above, therefore, concerns quality at equal algorithmic latency; only HT-Demucs is also deployable in real time on our hardware.
At $23$~ms, computation nearly doubles the algorithmic latency: one HT-Demucs pass takes about $20.6$~ms, so each separated sample leaves the pipeline about $44$~ms after it arrives.
Offline pipelines reach RTFs two orders of magnitude lower.
% \looseness=-1

\section{Conclusion}\label{sec:conclusion}
In this work, we showed that the algorithmic latency of an offline separator can be reduced to a single STFT hop by choosing where to read its output, and that the model's STFT window predicts where quality starts to drop sharply. 
The computational cost remains constant across the entire latency range.
Offline models can therefore run in a streaming regime without changing their architecture or weights.
At equal algorithmic latency, the public checkpoints of HT-Demucs and SCNet, streamed as released, reach a median SDR on par with published purpose-built real-time models, and HT-Demucs runs faster than real time on a GPU.
Finally, lookahead offers a tunable trade-off between latency and quality. 
Above one STFT window, gains remain below 1 dB for up to 3 s and depend mainly on the architecture; below that threshold, performance falls steeply as synthesis frames overlap the STFT padding.
% \looseness=-1

Purpose-built real-time models retain a large advantage in computational efficiency, which matters on embedded or edge hardware.
Emitting several output hops per forward pass, at the cost of added latency, and quantizing weights are promising directions for future work to reduce the cost of streamed models.

\newpage

\subsection*{Acknowledgements}
This project was provided with computing and storage resources by GENCI at IDRIS under grant 2025-AD011017041 on the supercomputer Jean Zay's A100 partition.
\\
\\
LLMs were used to help improve language and clarity during the editing process.

\subsection*{Compliance with Ethical Standards}
This work relies exclusively on numerical simulations conducted on publicly available, open-access datasets, for which no ethical approval was required.

% References should be produced using the bibtex program from suitable
% BiBTeX files (here: strings, refs, manuals). The IEEEbib.bst bibliography
% style file from IEEE produces unsorted bibliography list.
% -------------------------------------------------------------------------
\bibliographystyle{IEEEbib}
\bibliography{MELODIA}

\end{document}